# Cantilever-based electret energy harvesters

**S Boisseau[1], G Despesse[1], T Ricart[1], E Defay[1] and A Sylvestre[2]**

[1] CEA/LETI, 17 avenue des martyrs, Minatec Campus, Grenoble, France
[2] G2ELab, CNRS, 25 avenue des Martyrs, Grenoble, France

E-mail: sboisseau@gmail.com

**Abstract.** Integration of structures and functions has permitted to reduce electric consumptions of sensors, actuators and electronic devices. Therefore, it is now possible to imagine low-consumption devices able to harvest energy in their surrounding environment. One way to proceed is to develop converters able to turn mechanical energy, such as vibrations, into electricity: this paper focuses on electrostatic converters using electrets. We develop an accurate analytical model of a simple but efficient cantilever-based electret energy harvester. Therefore, we prove that with vibrations of 0.1g (~1m/s²), it is theoretically possible to harvest up to 30µW per gram of mobile mass. This power corresponds to the maximum output power of a resonant energy harvester according to the model of William and Yates. Simulations results are validated by experimental measurements, raising at the same time the large impact of parasitic capacitances on the output power. Therefore, we 'only' managed to harvest 10µW per gram of mobile mass, but according to our factor of merit, this puts us in the best results of the state of the art.

## 1. Introduction

Thanks to size reduction, micro-electro-mechanical-systems (*MEMS*) are consuming less and less energy, giving them the opportunity to harvest energy in their surrounding environment. This field of research called 'Energy Harvesting' consists in the development of converters able to turn ambient energy (light, air flux, variation of temperatures, vibrations) into electricity that is used to power the microsystem. Many principles of conversion have already been developed: photovoltaic, thermoelectric, biofuelcells... As for mechanical energy from vibrations, it can be converted by three main different principles: piezoelectric, electromagnetic and electrostatic conversion. In this study, we focus on electrostatic converters which are based on a capacitive architecture (two charged electrodes spaced by an air gap) and connected to a load. Vibrations induce changes in the geometry of the capacitor and a circulation of charges between electrodes through the electrical load. The electronic circuit that manages power conversion of 'standard' electrostatic energy harvesters [1] is quite complicated and induces losses and therefore a decrease of efficiency. To limit the use of a management electronic circuit, it is possible to use electrets (stable electrically charged dielectrics) that polarize the capacitance and allow to harvest energy from vibrations without using cycles of charging and discharging.

Many electret-based energy harvesters have been developed and have proven the interest of such devices [2-22]. Most of these devices are in-plane structures where the variation of capacitance is obtained by a variation of surface between patterned electrodes, while the gap is kept constant. These structures are generally hard to manufacture using elaborate clean room processes and especially DRIE (Deep Reactive Ion Etching) but give the opportunity to harvest energy when the vibrations of the ambient environment are not constant because they avoid contacts between electrets and electrodes. In this paper, we have chosen to study a simpler structure: the 'cantilever-based electret energy harvester'. This structure does not maximize the output power of the energy harvester if vibrations are not well defined but is particularly suitable when vibrations are stable in terms of frequency and amplitude in the time. Moreover, this structure is quite easy to manufacture and therefore low-cost.

In section 2, we present the theory of vibration energy harvesting and more especially of energy harvesting using electrets. Then, we develop an accurate analytical model, its implementation (under Matlab/Simulink) and its validation using FEM (Finite Element Method). Thanks to this model, the structure can be optimized, as presented in section 4. Finally, in section 5, we present experimental results and a comparison to simulation results. We finally develop a model taking parasitic capacitances into account to explain the differences between our first model and experimental results.

## 2. Electret-based energy harvesters using cantilevers

Our electret-based energy harvester is a microsystem able to convert mechanical energy from vibrations into electricity. It is part of vibration energy harvesters whose general model is presented hereafter.

*2.1. William and Yates' general model for vibration energy harvesters*

Regardless of the conversion principle (electrostatic, electromagnetic or piezoelectric), resonant energy harvesters can be modeled as a mobile mass ($m$) suspended to a support by a spring ($k$) and damped by forces ($f_{elec}$ and $f_{mec}$). When a vibration occurs $y(t) = Y \sin(\omega t)$, it induces a relative displacement of the mobile mass $x(t) = X \sin(\omega t + \varphi)$ compared to the frame (figure 1). Part of the kinetic energy of the moving mass is lost due to mechanical damping ($f_{mec}$) while the other part is converted into electricity, which is modeled by an electrostatic force ($f_{elec}$) in electrostatic energy harvesters. Ambient vibrations are generally low in amplitude (typically $Y=25\mu m$) and the use of a mass-spring structure enables to take advantage of a resonance phenomenon that amplifies the amplitude of vibrations perceived by the mobile mass and the harvested energy. Newton's second law gives the differential equation that rules the movement of the mobile mass (1).

$$m\ddot{x} + kx + f_{elec} + f_{mec} = -m\ddot{y} \tag{1}$$

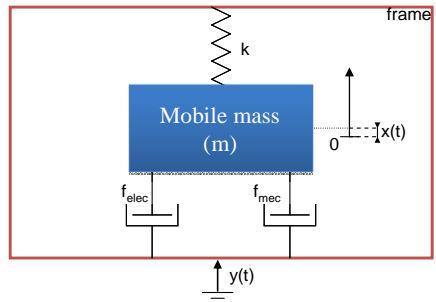

**Figure 1.** Mechanical system.

When forces can be modeled as viscous forces, $f_{elec} = b_e \dot{x}$ and $f_{mec} = b_m \dot{x}$, where $b_e$ and $b_m$ are respectively electrical and mechanical damping coefficients, William and Yates [23] have proven that the maximum output power of a resonant energy harvester subjected to an ambient vibration is reached when the natural angular frequency ($\omega_n$) of the mass-spring structure is tuned to the angular frequency of ambient vibrations ($\omega$) and when the damping rate $\xi_e = b_e/(2m\omega_n)$ of the electrostatic force $f_{elec}$ is equal to the damping rate $\xi_m = b_m/(2m\omega_n)$ of the mechanical friction force $f_{mec}$. This maximum output power $P_{W\&Y}$ can be simply expressed with (2), when $\xi_e = \xi_m = \xi$.

$$P_{W\&Y} = \frac{mY^2\omega_n^3}{16\xi} \quad (2)$$

As $P_{W\&Y}$ is a good approximation to know the output power of vibration energy harvesters when forces are modeled as viscous forces, comparing the output power ($P$) of a resonant energy harvester to $P_{W\&Y}$ gives a legitimate factor of merit $\alpha_{W\&Y}$:

$$\alpha_{W\&Y} = \frac{P}{P_{W\&Y}} \quad (3)$$

Nevertheless, in many studies, the weight of the mobile mass is not given while the surface area of the electrodes ($S$) is often provided. Therefore, to compare systems, we had developed, in a previous study, an other factor of merit, normalized by the active surface $S$ in place of the mass [24]:

$$\chi = \frac{P}{Y^2\omega^3 S} \quad (4)$$

These two factors of merit will be used in the next parts, to compare our system to the state of the art.

*2.2. Cantilever-based electret energy harvesters – Principles and Model*

The particularity of electret-based energy harvesters is the use of an electret to maintain the electrostatic converter charged through time. Electrets are dielectrics able to keep an electric field (and a surface voltage $V$) for years thanks to charge trapping (figure 2). These materials are in electrostatics, the equivalent to magnets in magnetostatics.

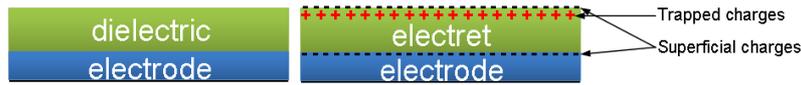

**Figure 2.** Electret film.

Electrets are obtained by implanting electric charges into dielectrics. Theoretically, dielectrics do not conduct electricity; therefore, the implanted charges stay trapped inside. Many techniques exist to manufacture electrets [25, 26], but the most common is the corona discharge (figure 3). It consists in a point-grid-plane structure whose point is subjected to a strong electric field: it leads to the creation of a plasma, made of ions. These ions are projected onto the surface of the sample to charge, and transfer their charges to its surface. This mechanism results in the implantation of charges at the surface (figure 2), into the bulk or at the interfaces of the material. The grid is used to limit the surface voltage of the electret to a wanted final value.

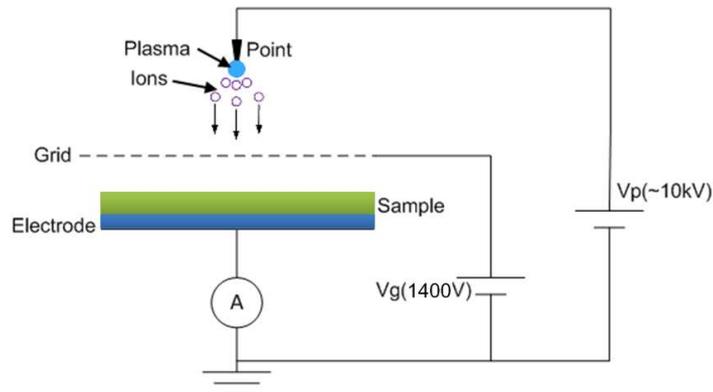

**Figure 3.** Corona discharge used to charge our device.

Nevertheless, dielectrics are not perfect insulators and implanted charges can move inside the material or can be compensated by other charges or environmental conditions, and finally disappear. A focal area of research on electrets concerns their stability [15, 27, 28]. Nowadays, many materials are known as good electrets able to keep their charges for years: for example, Teflon® and silicon dioxide ($SiO_2$) whose stability is estimated to more than 100 years [29-32].

The structure able to turn vibrations into electricity using electrets is introduced in figure 4: the system is composed of a counter-electrode and an electrode on which is deposited an electret, spaced by an air gap and connected by an electrical load (here a resistor). The electret has a constant charge $Q_i$, and, due to electrostatic induction and charges conservation, the sum of charges on the electrode and on the counter-electrode equals the charge on the electret: $Q_i = Q_1 + Q_2$.

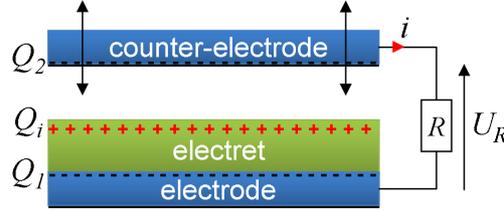

**Figure 4.** Electrostatic converter using electret.

When a vibration occurs, it induces a change in the capacitor geometry (*e.g.* the counter-electrode moves away from the electret, changing the air gap and then the electret influence on the counter-electrode) and a reorganization of charges between the electrode and the counter-electrode through the load. This induces a current across the load $R$ and part of the mechanical energy is then turned into electricity.

The converter introduced in figure 4 is integrated into a clamped-free beam mechanical structure (figure 5). The lower face of the beam is metalized and is used as the counter-electrode. The electret and the electrode are placed under the beam, separated by an air gap and electrically connected by a load (figure 5(a)). According to William and Yates' formula (2), the output power is proportional to the mobile mass; consequently, to increase the output power, a proof mass $m$ is added at the free end of the cantilever. Vibrations $y(t)$ induce a relative displacement $x(t)$ of the mobile mass compared to the electrode. Structure parameters are presented in figure 5(b): $h$ is the beam thickness, $w$ its width, $L$ the length between the clamping and the gravity center of the mass, $2L_m$ the mass length, $g(t)$ the thickness of the air gap between the counter-electrode and the electret, $g_0$ the thickness of this air gap without vibrations and without the weight effects $\vec{W}$, $V$ the surface voltage of the electret, $C_1$ the capacitance of the electret, $C_2$ the capacitance of the air gap and finally $\lambda$ the electrode length.

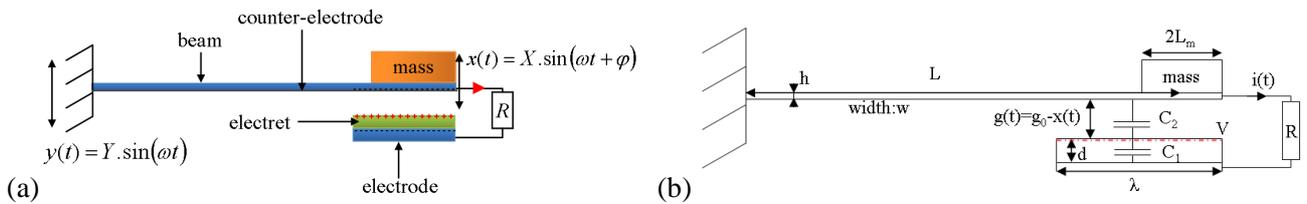

**Figure 5.** (a) Cantilever-based energy harvester using electrets. (b) Energy harvester parameters.

To identify the main parameters of this kind of energy harvesters and to maximize the output power for a given vibration ($y(t) = Y.\sin(\omega t)$), it is necessary to find coupled mechanical and electrostatic equations that rule the energy harvester.

## 3. Analytical model of the 'Cantilever-based electret energy harvester'

To determine the output power of the energy harvester for a given vibration $y(t)$, it is necessary to solve the equation of motion and to find the quantity of charge transferred to the output. Therefore, the goal of section 3 is to develop the analytical model of the 'cantilever-based electret energy harvester' parameterized in figure 5 for mechanical and electrostatic parts.

### 3.1. Model of the mechanical system

The clamped-free beam with a mass at the free end can be modeled as a damped mass-spring structure as presented in figure 1 and by adding the effect of weight $\vec{W} = m\vec{g}$. The mechanical friction forces can be modeled as viscous forces ($f_{mec} = b_m \dot{x}$) and the electrostatic force is the derivative of the electrostatic energy of the capacitor $W_e$ with respect to the displacement $x$. $W_e$ is equal to the charge on the upper electrode $Q_2$ squared, divided by two times the capacitance as a function of time $C(t)$. Thereby, the mechanical system is ruled by (5).

$$m\ddot{x} + b_m \dot{x} + kx - \frac{d}{dx}(W_e) - mg = -m\ddot{y} \Rightarrow m\ddot{x} + b_m \dot{x} + kx - \frac{d}{dx}\left(\frac{Q_2^2}{2C(t)}\right) - mg = -m\ddot{y} \quad (5)$$

To maximize the output power of the energy harvester, the natural angular frequency ($\omega_n = \sqrt{k/m}$) of the mass-spring structure has to be tuned to the angular frequency of the ambient vibrations ($\omega$). Moreover, according to equations from mechanical structures theory, the spring constant $k$ can be deduced from the beam geometric parameters as follow:

$$k = m\omega_n^2 = \frac{3EI}{L^3} = \frac{Ewh^3}{4L^3} \quad (6)$$

Where $E$ is the Young's modulus and $I$ the quadratic moment of the beam.

Because of the mass, the behavior of the beam has to be studied on two parts. A drawing of the structure is presented in figure 6 and shows the deformation of the cantilever $\delta(z)$ as a function of the position on the cantilever $z$ for a forced deflection $x$ at $z=L$. The first part ($z \in [0, L_1 = L - L_m]$) does not have an additional mass: its behavior corresponds to the one of a clamped-free beam whose deflection at the end ($x_1$) is imposed and given by $\delta(z) = \frac{x_1}{2L_1^3} z^2 (3L_1 - z)$. The second part that has the additional mass ($z \in [L_1, L_2 = L + L_m]$) follows the deflection of part 1: the derivative of the deflection ($\delta(z)$) with respect to the position ($z$) for part 2 is constant and equal to the derivative of the deflection of part 1 at $z=L_1$ (7).

$$c = \left.\frac{d\delta(z)}{dz}\right|_{z=L_1} = \left.\frac{d\delta(z)}{dz}\right|_{z \in [L_1, L_2]} = \frac{3}{2}\frac{x_1}{L_1} \text{ with } x_1 = x - cL_m \quad (7)$$

Therefore, for a given static deflection ($x$) on the position $L$ of the beam, the deformation of the beam can be simply expressed as a function of the parameters in both parts:

$$\begin{cases} \delta(z) = \frac{x_1}{2L_1^3} z^2 (3L_1 - z) & \text{[part 1]} \\ \delta(z) = c(z - L) + x & \text{[part 2]} \end{cases} \quad (8)$$

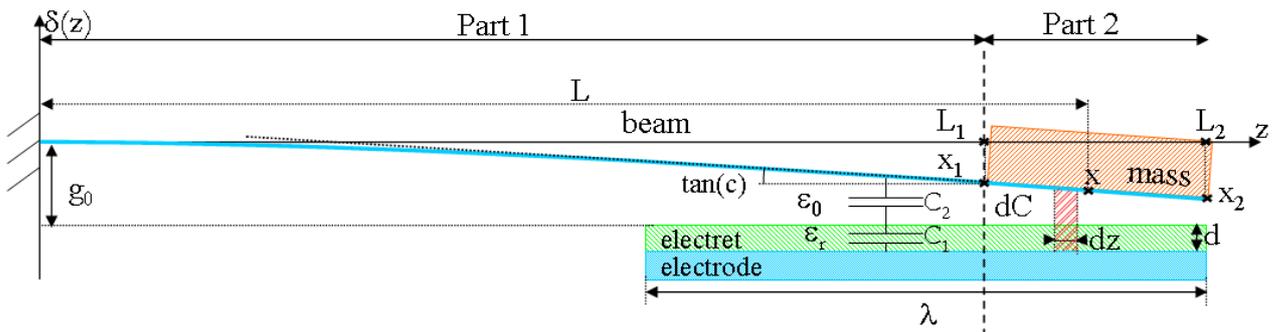

**Figure 6.** Deformation of the cantilever for an imposed deflection ($x$).

Figure 7 presents the beam deformation resulting from equation (8) for a beam of $L$=30mm and $L_m$=2mm and for an imposed static displacement of $x$=300µm compared to the deformation performed by FEM calculation (Comsol® Multiphysics). It proves that our calculations fit with FEM results.

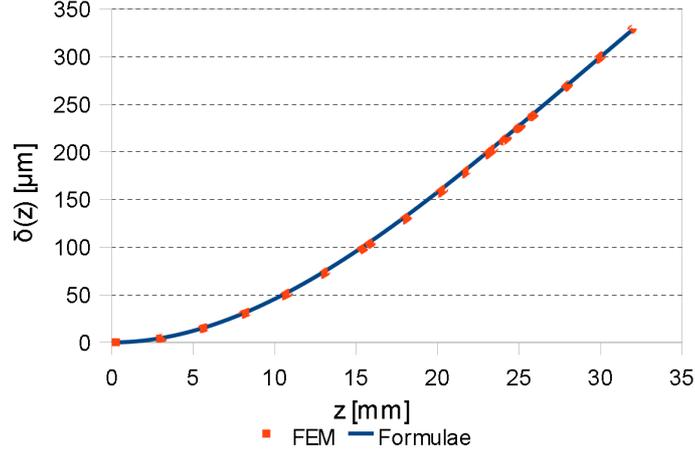
**Figure 7.** Deformation of the cantilever for an imposed static deflection (*x*).

Nevertheless, the problem we want to solve is not static but dynamic. Therefore, it is useful to verify that the beam deformation behavior is the same in dynamic and in static. We have verified this using FEM : it confirms that the deformation in dynamic and in static can be considered as equivalent. Thus, we can consider that the deflection in dynamics can be simply expressed with (8) assuming that *x* is the imposed deflection on the mass gravity point.

*3.2. Modeling of the electrostatic system*

The equivalent model of the energy harvester is presented in figure 8, where $Q_2$ is the charge on the counter-electrode, *V* the surface voltage of the electret and *C(t)* the capacitance between the beam and the electrode. This capacitance corresponds to the serial capacitance formed by the electret dielectric material capacitance $C_1$ and the air gap capacitance $C_2(t)$. Kirchhoff's laws give the differential equation that governs the electrostatic system (9):

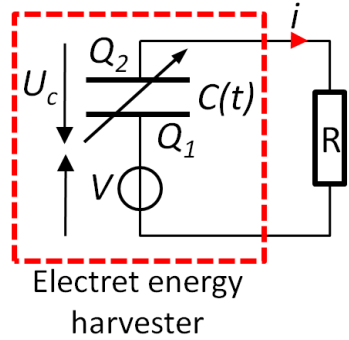

Electret energy harvester

**Figure 8.** Equivalent electric model of the energy harvester.

$$\frac{dQ_2}{dt} = \frac{V}{R} - \frac{Q_2}{R} \times \left[\frac{1}{C(t)}\right] = \frac{V}{R} - \frac{Q_2}{R} \times \left[\frac{1}{C_1} + \frac{1}{C_2(t)}\right] \quad (9)$$

Moreover, the electrostatic energy stored in the capacitor is:

$$W_e = \frac{1}{2}\frac{Q_2^2(t)}{C(t)} \quad (10)$$

To solve (9), it is necessary to know the capacitance of the electrostatic converter as a function of the imposed deflection (*x*). Knowing the cantilever deformation, and considering a capacitor of infinitesimal length (*dz*) (figure 6), one can get the infinitesimal capacitance on both part (*dC$_{p1}$* and *dC$_{p2}$*) for a given *x*.

$$\begin{cases} dC_{p_1}(x) = \dfrac{\varepsilon_0 w.dz}{g_0 - \delta(z) + \dfrac{d}{\varepsilon_r}} & \text{with } \delta(z) = \dfrac{x_1}{2L_1^3} z^2(3L_1 - z) \quad \text{[part 1]} \\ dC_{p_2}(x) = \dfrac{\varepsilon_0 w.dz}{g_0 - \delta(z) + \dfrac{d}{\varepsilon_r}} & \text{with } \delta(z) = c(z - L) + x \quad \text{[part 2]} \end{cases}$$ (11)

By integrating these expressions, the total capacitance between both electrodes is:

$$C(x) = C_{p_1}(x) + C_{p_2}(x) = \varepsilon_0 w \int_{L_2-\lambda}^{L_1} \dfrac{dz}{g_0 - x\dfrac{z^2(3L-z)}{2L^3} + \dfrac{d}{\varepsilon_r}} + \dfrac{\varepsilon_0 w}{c} \ln\left( \dfrac{g_0 + \dfrac{d}{\varepsilon_r} + cL_m - x}{g_0 + \dfrac{d}{\varepsilon_r} - cL_m - x} \right)$$ (12)

The integral defining $C_{p_1}(x)$ cannot be analytically calculated and will be numerically computed.

This capacitance expression has been compared to a FEM simulation and the curves presented in figure 9 show that results are in excellent agreement. These results were also compared to the formula of a simple plane capacitor neglecting fringe effects ($C(x) = \varepsilon_0 S/(g_0 - x + d/\varepsilon_r)$), where $S$ is the surface of the electrodes, $g_0$ the initial gap and $x$ the imposed deflection. With our parameters ($L_m$=2mm, $L$=30mm, $g_0$=505µm, $d$=100µm, $w$=12.33mm, $\varepsilon_r$=2, $\lambda$=10mm), we have found that the model of the simple plane capacitor overestimate (up to 35%) the maximal capacitance of the energy harvester.

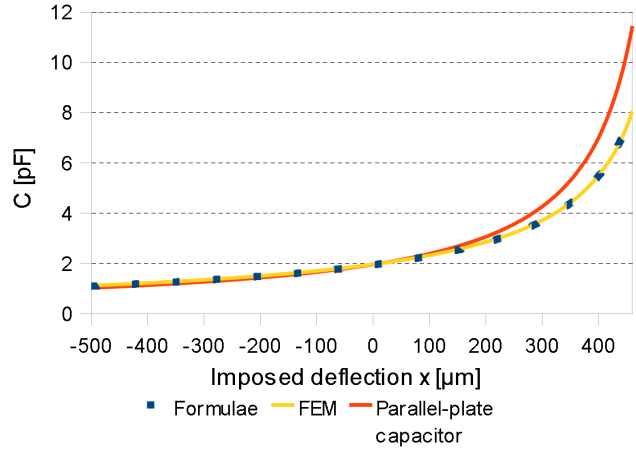

**Figure 9.** Capacitance between the electrodes (*C*) versus forced displacement (*x*) ($L_m$=2mm, $L$=30mm, $g_0$=505µm, $d$=100µm, $w$=12.33mm, $\varepsilon_r$=2, $\lambda$=10mm).

This accurate value of the capacitance for a given deflection is then applied in the mechanical system introduced in section 3.1.

*3.3. Complete analytical model*

In order to get the output power of the energy harvester, mechanical and electrostatic systems have to be coupled. From (5) and (9), one can find that the system of equations that governs the energy harvester is (13).

$$\begin{cases} m\ddot{x} + b_m.\dot{x} + kx - \dfrac{d}{dx}\left(\dfrac{Q_2^2}{2C(t)}\right) - mg = -m\ddot{y} \\ \dfrac{dQ_2}{dt} = \dfrac{V}{R} - \dfrac{Q_2}{C(t)R} \end{cases}$$ (13)

Nevertheless, it is not possible to get an analytic expression of *x* and $Q_2$. Therefore, the system is numerically solved in Matlab/Simulink (figure 10).

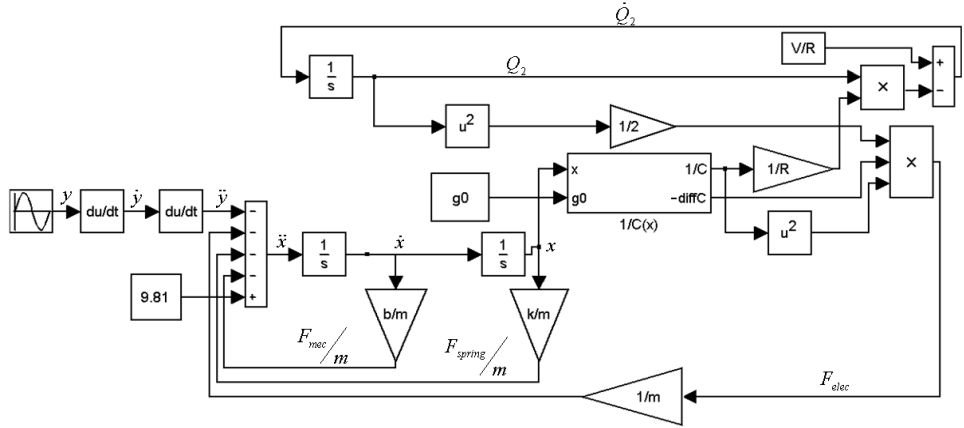

**Figure 10.** Simulink model of the energy harvester.

The deflection (*x*) given by Matlab and the voltage across the resistor $U_R = R(dQ_2/dt)$ versus time are presented in figure 11(a) and 11(b) for *V*=1400V, *d*=127µm, $g_0$=1mm, $\lambda$=20mm, *R*=300MΩ.

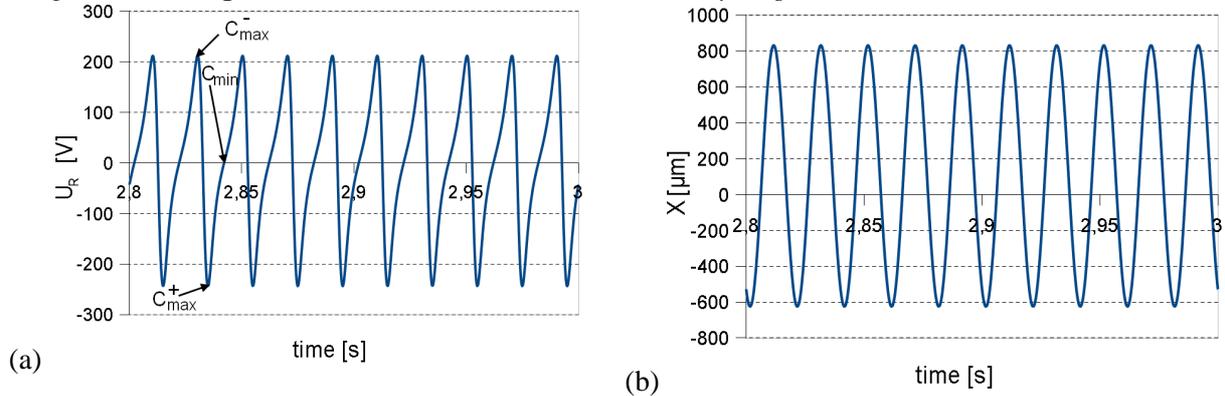

**Figure 11.** (a) Example of output voltage and (b) deflection versus time.

Figure 11(a) shows that the output voltage of 'cantilever-based electret energy harvesters' can be higher than 200V. This can greatly simplify rectification of the output voltage using diode bridges. Moreover, figure 11(a) shows a particularity of the output voltage of cantilever-based electret energy harvesters: output voltage presents a discontinuity when it passes from its higher value (the capacitance is just before its maximum $C_{max}^-$) to its lower value (the capacitance is just after its maximum $C_{max}^+$) because the current changes direction when the capacitance crosses its maximum. The current also changes direction when the capacitance crosses its minimum $C_{min}$. But, since the output voltage equals 0 when the capacitance is minimum, no discontinuity on the output voltage appears.

In section 3, we have developed the complete analytical model of the energy harvester and its implementation on Simulink. Thanks to this model, the system can be optimized to give the maximum output power.

## 4. Output power and Optimization

The goal of this section is to maximize the average output power of the energy harvester *P* for a given vibration *y(t)*. Actually, the power can be simply computed from the derivative of $Q_2$ given by the Simulink model presented in figure 10 (14). We will determine in section 4.1 the parameters to optimize before optimizing them in section 4.2.

$$P = \frac{1}{t_2 - t_1} \int_{t_1}^{t_2} R \left( \frac{dQ_2}{dt} \right)^2 dt \quad \text{where } t_1 \text{ and } t_2 \text{ are times taken in the steady state} \tag{14}$$

*4.1. Parameters to optimize*

We have chosen to consider $L$, $L_m$, $m$, $w$, $k$, $Y$, $\omega=2\pi f$, $V$, $\varepsilon_r$, $d$ and $\xi$ as given parameters and the load $R$, the electrode length $\lambda$ and the initial air gap $g_0$ as parameters to optimize. Figures 12(a), 12(b) and 12(c) present the output power of the energy harvester when one parameter varies while the other are kept constant. Constant values are: $Y=10\mu m$, $f=50Hz$, $m=5g$, $\xi=1/150$, $V=1400$, $w=12.3mm$, $d=127\mu m$, $\varepsilon_r=2$, $L_m=2mm$, $L=30mm$ and $R=1G\Omega$, $\lambda=10mm$, $g_0=2mm$ when one of them varies.

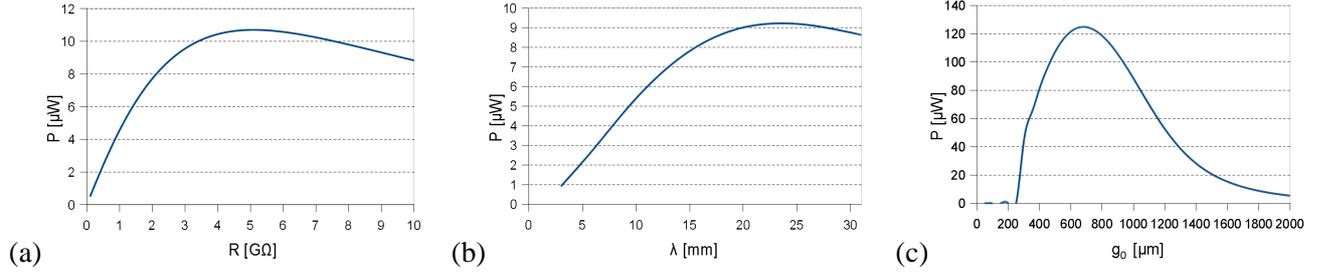

**Figure 12.** Parameters to optimize and their effect on the average output power: (a) load, (b) length of the electrode and (c) initial gap.

It is obvious that those three parameters play an important role in the behavior of the energy harvester and it is also obvious that an optimum is present in each curve signifying that optimal parameters exist. As these parameters are not independent, they have to be optimized together in a single loop.

*4.2. Optimization and maximum output power*

The optimization process uses the *fminsearch* Matlab function to minimize the inverse of the output power ($1/P$) considering ($R$, $\lambda$ and $g_0$) as parameters. Table 1 gives the maximum output powers and the optimal parameters for different mobile masses for an 'ambient' vibration of $(Y, f)=(10\mu m, 50Hz)\approx 1ms^{-2}$: it proves that the model of William and Yates gives a good approximation of the output power but is not rigorously exact in 'cantilever-based electret energy harvesters'. Actually, if the system could be modeled by William and Yates model, $\alpha_{W\&Y}$ should be equal to *1* whatever the value of the mobile mass. Moreover, when the electrostatic force is sufficiently high, which is directly linked to the surface voltage of the electret, output power of the energy harvester is bigger than the output power determined by William and Yates' model. When the surface voltage of the electret is not high enough to induce a sufficient electrostatic force that can absorb the kinetic energy of the mobile mass, it cannot permit to obtain the optimal energy with the mobile mass (e.g. when $m=10g$). For $m=5g$, a surface voltage of 1400V should allow to harvest 160µW which corresponds to a power density per mass unit of ~30µW/g.

**Table 1.** Output Power ($P$) as a function of the mass ($m$) with V=1400V.

| $m$ (g) | $R_{opt}$ (GΩ) | $\lambda_{opt}$ (mm) | $g_{0opt}$ (µm) | $P$(µW) | $P_{W\&Y}$ (µW) | $P/m$ (µW/g) | $\alpha_{W\&Y}$ |
|---|---|---|---|---|---|---|---|
| 1 | 10 | 6.4 | 700 | 36.59 | 29.07 | 36.59 | 1.26 |
| 2 | 7.1 | 5.9 | 602 | 71.7 | 58.14 | 35.85 | 1.23 |
| 3 | 4.36 | 7.4 | 600 | 104 | 87.21 | 34.67 | 1.19 |
| 5 | 2.18 | 9.6 | 593 | 160 | 145.34 | 32 | 1.1 |
| 10 | 0.8 | 14.30 | 901 | 173 | 290.68 | 17.3 | 0.6 |

The results presented in table 1 are given with an accuracy of 1µm for $g_0$ and 10µm for $\lambda$. These precisions will not be easy to obtain. To see the effect of inaccuracies on $\lambda$ and $g_0$ on the response of the system, we have plotted the output power of the system when $g_0$ and $\lambda$ range from 100µm to their optimal values (0 corresponds to the optimal value). Results presented in figure 13 prove that the output power does not vary much near the optimal values of $g_0$ and $\lambda$, but to avoid a contact between the counter-electrode and the electret, for the prototype, we will choose a value of $g_0$ slightly higher than the optimal value. Therefore, even with inaccuracies on $\lambda$ and $g_0$ (~50µm), the output power should be equal to at least 140µW.

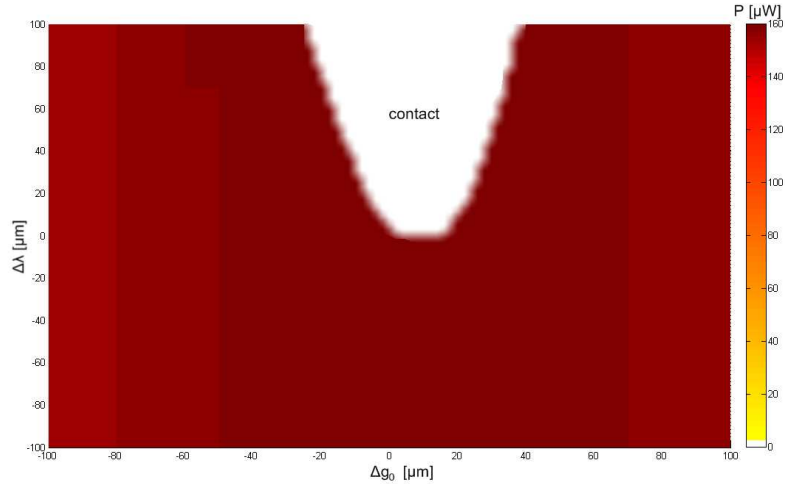

**Figure 13.** Output power in function of $\lambda$ and $g_0$ variations around their optimal value.

Similarly, the effect of the frequency and the amplitude of vibrations were evaluated on the optimal design. As the system is resonant and low damped, $f$ and $Y$ variations induce a large output power change (figure 14): these parameters are critical but can be adjusted with a good accuracy. Therefore, it proves that, when the parameters of vibrations are constant, cantilever-based electret energy harvesters are good energy harvesters. But, if the amplitude of vibrations increases, it can lead to a contact between the upper electrode and the electret that can damage this latter.

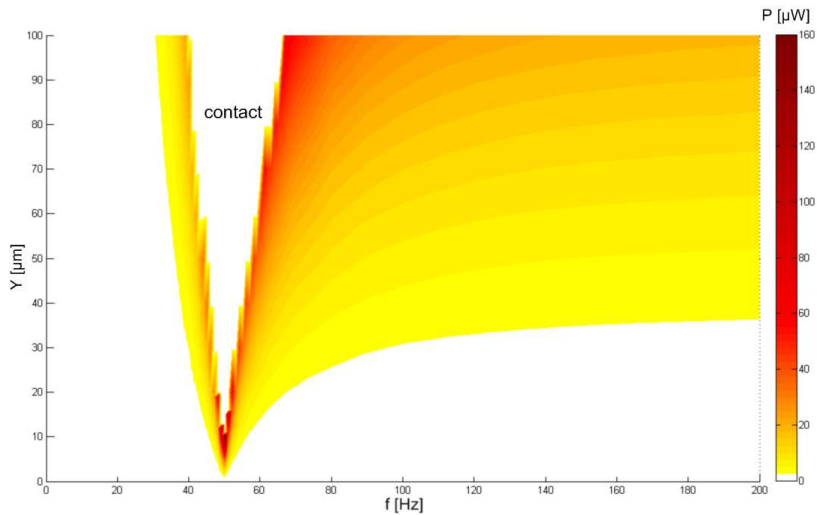

**Figure 14.** Effect of the vibrations on the optimized system.

The resonant system has been optimized and theoretical results have proven that up to 160µW could be reached with low vibrations (10µm@50Hz)≈1ms$^{-2}$. These parameters are now tested on a prototype.

## 5. Experimental results and new model taking parasitic capacitances into account

The goal of experimental results presented in section 5 is to validate theoretical results that have been obtained in the previous sections and to see the limits of our model. We conclude this section with a better analytical model of the energy harvester that takes parasitic capacitances into account.

## 5.1. Prototype and expected output power

To insure a good flatness, the beam is made in silicon and attached to the frame at one end. The mobile mass attached to the other end is made in tungsten that has a high density (d=17) to limit its size. The electret is made in Teflon FEP (Fluorinated Ethylene Propylene) that is well known for being a good electret [32]. In this study, the electret is charged to 1400V. Our prototype design is presented in figure 15.

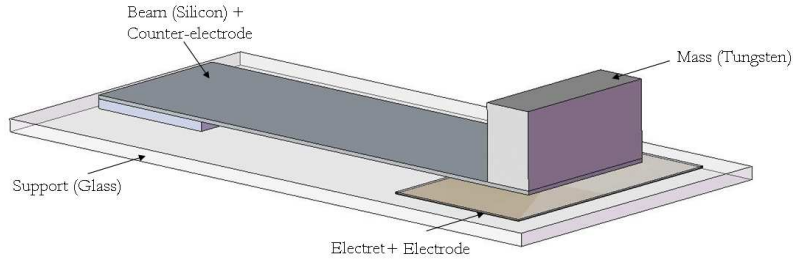

**Figure 15.** Cantilever-based electret energy harvester – Schema.

The natural frequency of the structure, computed with FEM, is actually 48.77Hz. The difference with our model is due to the simplification in (6). To fit the vibrations imposed by the environment (50Hz), the beam width is slightly changed to 13mm. Therefore, the output power of the energy harvester should be 140µW and the voltage across the resistance versus time should look like the one presented on figure 16(a) given by simulation (figure 16(b) is a zoom of figure 16(a)).

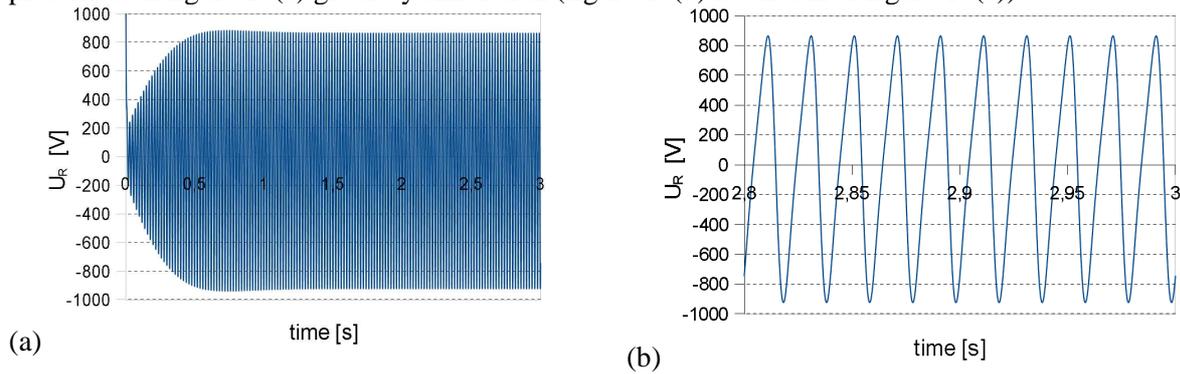

**Figure 16.** (a) Output voltage of the energy harvester versus time. (b) Zoom of figure 15(a) ($m$=5g, $V$=1400V, $R$=2.18GΩ, $\lambda$=9.6mm, $g_0$=593µm, $Y$=10µm, $f$=50Hz).

The prototype was made on a glass support (figure 17(a)) to limit parasitic capacitances. A mass of 5g in tungsten was added at the free end of the cantilever. The electret is obtained by evaporating a 300nm-thick layer of aluminum on the rear face of a Teflon FEP film. It is glued on a sheet of copper to ensure the flatness of the electret during charging. The electret is charged using a standard corona discharge as presented in figure 3 with a point voltage $V_p$ of 10kV and a grid voltage $V_g$ of 1400V. It is placed in an oven at 175°C and cooled to the ambient temperature while charging to improve stability. The long-term stability was not studied but the short-term (some days) experiments showed low charge losses (-0.21% in 8 days) (figure 17(b)).

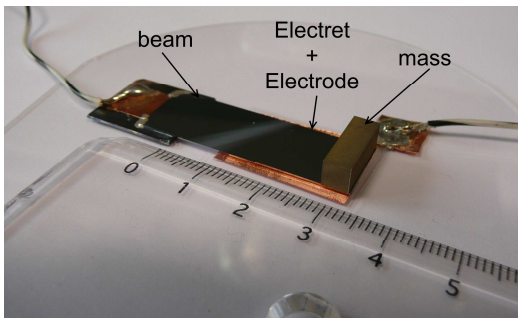
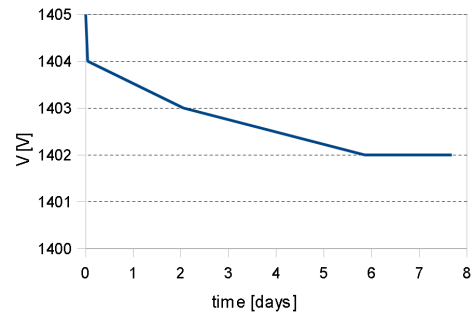

**Figure 17.** (a)Cantilever-based electret energy harvester – prototype. (b) Stability of Teflon FEP electret charged at 1400V

*5.2. Output Power, comparison to the theory and limits*

We present hereafter the experimental results we got on our prototype. The optimized parameters were applied to the prototype introduced in figure 17(a). The output power of our prototype is presented in figure 18(a). Experimental and theoretical curves do not fit and the output power is much lower than expected. These differences are due to parasitic capacitances that become important when using loads of high values. In those cases, the model given in figures 5(b) and 8 should be modified to take a parasitic capacitance $C_{par}$ in parallel with the load into account.

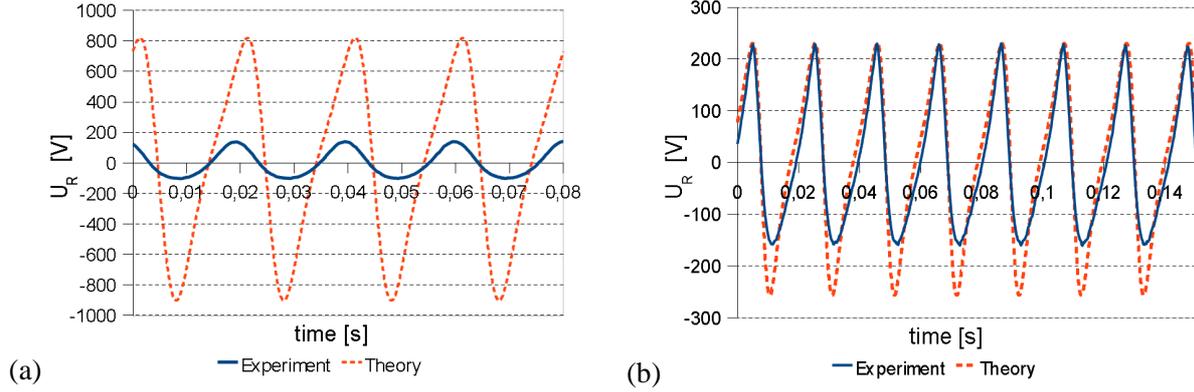

**Figure 18.** Experimental output voltages (a) for R=2,2GΩ and (b) for R=300MΩ.

To avoid these phenomena, the value of the load was constraint to 300MΩ (chosen after some experimental measurements in order to limit the parasitic capacitances induced by the load) and the optimization process was restarted on $g_0$ and $\lambda$. 'New' optimized values are presented in table 2.

**Table 2.** Parameters and values.

| Parameter | Designation | Value |
|---|---|---|
| $M_{beam}$ | Material of the beam | Silicon |
| E | Young's Modulus of Silicon | 160 GPa |
| L | distance between the clamping and the centre of gravity of the mass | 30 mm |
| h | Thickness of the beam | 300 µm |
| w | Width of the beam / Width of the electret | 13 mm |
| $2L_m$ | Length of the mobile mass | 4 mm |
| m | Mobile mass | 5 g |
| $\omega_n = \omega$ | Natural angular frequency/Angular frequency of vibrations | $50 \times 2\pi$ rad/s |
| $Q_m = (2\xi_m)^{-1}$ | Mechanical quality factor of the structure | 75 |
| $M_{electret}$ | Material of the electret | FEP |
| $\varepsilon_r$ | Dielectric constant of the electret | 2 |
| d | Thickness of the electret | 127 µm |
| V | Surface voltage of the electret | 1400 V |
| $g_0$ | Thickness of the initial air gap | 700 µm |
| $\lambda$ | Length of the electrode | 22.8 mm |
| R | Load | 300 MΩ |

The structure was tested again with the new load and its output voltage is presented in figure 18(b). Experimental and theoretical curves fit, except for negative voltages. This is once again due to parasitic capacitances that clip the signal in its negative part. The mean output power of the energy harvester is 50µW when it is submitted to vibrations of 10µm@50Hz (1ms$^{-2}$) (our simulation predicted 80µW).

**Table 3.** Comparison to 7 prototypes among the most recent electret energy harvesters in the state of the art.

| Author | Ref | Vibrations | Active Surface (S) | Electret Potential (V) | Output Power (P) | Figure of merit ($\chi$) |
|---|---|---|---|---|---|---|
| Suzuki | [14] | 1mm@37Hz (54.0ms$^{-2}$) | 2.33 cm² | 450V | 0.28µW | 9.56×10$^{-5}$ |
| Halvorsen | [16] | 2.8µm@596Hz (39.2ms$^{-2}$) | 0.48 cm² | | 1µW | 5.06×10$^{-2}$ |
| Kloub | [17] | 0.08µm@1740Hz (9.6ms$^{-2}$) | 0.42 cm² | 25V | 5µW | 14.2 |
| Naruse | [18] | 25mm@2Hz (3.9ms$^{-2}$) | 9 cm² | | 40µW | 3.58×10$^{-2}$ |
| Edamoto | [19] | 500µm@21Hz (8.7ms$^{-2}$) | 3 cm² | 600 V | 12µW | 6.97×10$^{-2}$ |
| Miki | [20] | 100µm@63Hz (15.7ms$^{-2}$) | 3 cm² | 180V | 1µW | 5.37×10$^{-3}$ |
| Honzumi | [21] | 9.35µm@500Hz (92ms$^{-2}$) | 0.01 cm² | 52V | 90 pW | 3.32×10$^{-5}$ |
| This work (th.) | | 10µm@50Hz (1.0ms$^{-2}$) | 4.16cm² | 1400V | 152µW | 117.84 |
| This work (exp.) | | 10µm@50Hz (1.0ms$^{-2}$) | 4.16cm² | 1400V | 50µW | 38.75 |

Our experimental results correspond to a factor of merit $\alpha_{W\&Y}$ equals to 34% and to a factor of merit $\chi$ equals to 38.75, putting us in the best results of the state of the art (table 3), yet, our experimental results are quite different from theoretical results.

*5.3. Model taking parasitic capacitances into account*

In order to explain the differences between our theory and our experimental results, we have developed a new model that takes parasitic capacitances into account. The parasitic capacitance of the whole system is modeled as a capacitor $C_{par}$ in parallel with the energy harvester and the load, as presented in figure 19. $U$ is the voltage across the resistor, the parasitic capacitance and the electret energy harvester.

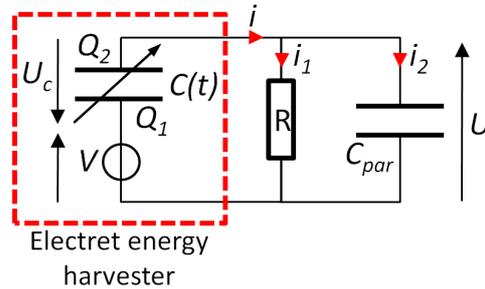

**Figure 19.** Equivalent electric model of the energy harvester taking parasitic capacitance into account.

In order to model the behavior of the energy harvester taking parasitic capacitances into account, the equation that rules the electrostatic part is modified as follow (15) (obtained using Kirchhoff's laws) while the equation that rules the mechanical part is the same as in (5) and (13).

$$\frac{dQ_2}{dt} = \frac{1}{\left(1+\dfrac{C_{par}}{C(t)}\right)}\left(\frac{V}{R} - Q_2\left(\frac{1}{RC(t)} - \frac{C_{par}}{C(t)^2}\frac{dC(t)}{dt}\right)\right) \qquad (15)$$

And the instantaneous harvested power is given by (16):

$$p(t) = \frac{U^2}{R} = \frac{1}{R}\left(V - \frac{Q_2}{C(t)}\right)^2 \qquad (16)$$

Our Simulink model was modified to take these changes into account. Our experimental results where then compared to our theoretical results taking parasitic capacitances into account (figures 20(a) and 20(b)) where parasitic capacitances with the 300MΩ load are estimated to 5pF and to 10pF with the 2.2GΩ load.

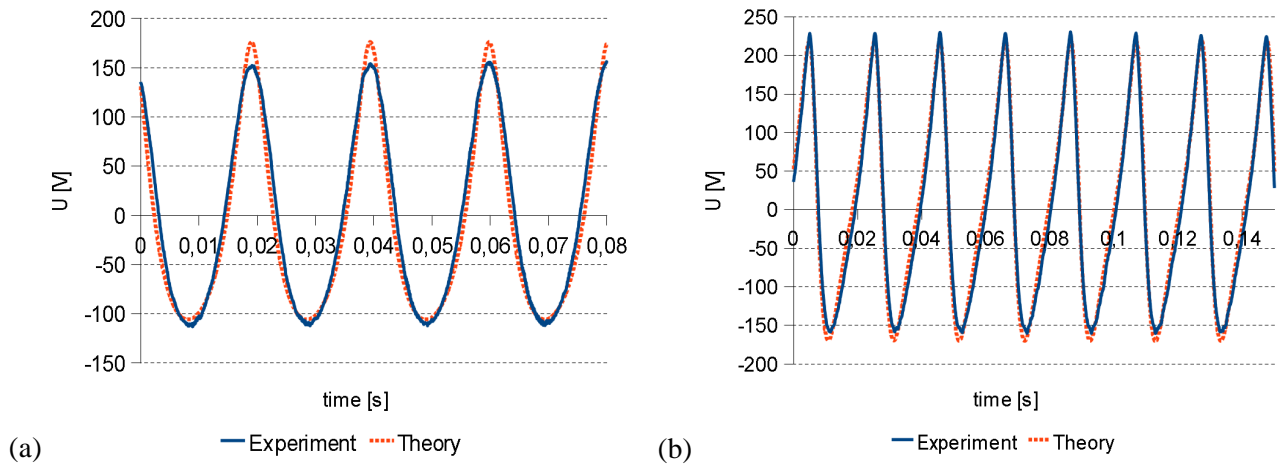

**Figure 20.** Experimental output voltages (a) for R=2,2GΩ and (b) for R=300MΩ and comparison to theory taking parasitic capacitances into account.

Figures 20(a) and 20(b) show that theoretical and experimental results fit perfectly and validate our new model. Therefore, it appears that parasitic capacitances have a large impact on the behavior of the energy harvester, decreasing the harvested power, especially when using high-value resistors. Unfortunately, as parasitic capacitances greatly depend on the load, restarting an optimization process taking parasitic capacitances into account would be difficult. Moreover, it would have a limited interest since parasitic capacitances can change a lot with the use of management electronic circuits. Therefore, to limit their effects, the load should be chosen so as not to exceed $Z_{par} = 1/C_{par}\omega$, the impedance of the parasitic capacitances which is roughly equal to $Z_{par}$=500MΩ in our case.

## 6. Conclusion and perspectives

We developed an analytical model of 'cantilever-based electret energy harvester' that is in agreement with FEM results. The optimization process has shown that the power harvested by these structures are in the same magnitude as theoretical output powers developed by William and Yates as soon as the surface voltage of the electret is sufficient to absorb the kinetic energy of the mobile mass. Finally, we validated our model with experimental results which reach up to 10μW per gram of mobile mass for low ambient vibrations of 0.1g (1ms$^{-2}$), using a resonant system.

Cantilever-based energy harvester can be a good low-cost solution to harvest energy when vibrations are constant in frequency and amplitude. The output power meets the magnitude of powers reached by piezoelectric or electromagnetic solutions.